\documentclass[a4paper,aps,prl,floats,twocolumn,superscriptaddress,showpacs,floatfix]{revtex4}

\usepackage{graphicx}
\usepackage{latexsym,amsmath,verbatim}

\newcommand{\avk}[1]{\langle {#1}\rangle}

\newcommand{\be}{\begin{equation}}
\newcommand{\ee}{\end{equation}}
\newcommand{\bea}{\begin{eqnarray}}
\newcommand{\eea}{\end{eqnarray}} 

\begin{document}

\title{Routes to thermodynamic limit on scale-free networks}

\author{Claudio Castellano}
\affiliation{SMC, INFM-CNR and
Dipartimento di Fisica, ``Sapienza'' Universit\`a di Roma,
P.le Aldo Moro 2, I-00185 Roma, Italy}

\author{Romualdo Pastor-Satorras}
\affiliation{Departament de F\'\i sica i Enginyeria Nuclear, Universitat
  Polit\`ecnica de Catalunya, Campus Nord B4, 08034 Barcelona, Spain}

\date{\today}

\begin{abstract}
  We show that there are two classes of finite size effects for
  dynamic models taking place on a scale-free topology.  Some models
  in finite networks show a behavior that depends only on the system
  size $N$.  Others present an additional distinct dependence on the
  upper cutoff $k_c$ of the degree distribution. Since the infinite
  network limit can be obtained by allowing $k_c$ to diverge with the
  system size in an arbitrary way, this result implies that there are
  different routes to the thermodynamic limit in scale-free networks.  The
  contact process (in its mean-field version) belongs to this second
  class and thus our results clarify the recent discrepancy between theory
  and simulations with different scaling of $k_c$ reported in
  the literature.
\end{abstract}


\maketitle

The thermodynamic limit is a crucial concept in statistical physics.
For example, no true singularity can occur in a system composed of a
finite number $N$ of elements at nonzero temperature
\cite{goldenfeld}, so that the very concept of phase-transition is
well defined only in the limit $N \to \infty$.  However, all real
systems are finite, and more so in numerical simulations, and to study
their properties one has to understand the role of finite size effects
when the limit $N \to \infty$ is taken.  In the case of critical
phenomena, the theory of finite size scaling (FSS)~\cite{privman90}
has successfully accomplished this task for processes on regular
lattices, allowing to detect the signature of continuous
phase-transitions even in very small systems. Finite size effects are
expected to be all the more relevant for systems with a strongly
heterogeneous interaction pattern, such as those described in terms of
complex networks \cite{barabasi02,mendesbook}.  Indeed, because of the
small-world property \cite{watts98} observed in most networks, the
number of neighbors that can be reached starting from a certain node
grows exponentially or faster with the geodesic distance. This implies
that, even for large networks, just a few steps are sufficient to
probe the finiteness of the system.  Moreover, most real networks
exhibit the scale-free (SF) property \cite{barabasi02}, i.e. they have
a degree distribution $P(k)$, defined as the probability that a vertex
is connected to other $k$ vertices, decaying for large $k$ as $P(k)
\sim k^{-\gamma}$ with $2 < \gamma \leq 3$, so that the local
topological properties display very strong fluctuations, increasing
with the size of the network \cite{dorogorev}.  In order to understand
phase transitions, and in general any kind of dynamical process, in SF
networks it is thus necessary to first comprehend how size effects,
and in particular FSS, work in this class of systems.

Some attempts in this direction have been already done.  For
equilibrium continuous phase transitions the situations seems to be
rather well established in some cases. For example, a FSS
phenomenological theory has been developed for the Ising model
\cite{hong07, dorogovtsev07:_critic_phenom}, based on a scaling ansatz for the
free energy, leading to the FSS form for the magnetization at zero
external field $m(\Delta, N) = N^{-\beta/\bar{\nu}}f(\Delta N^{1/\bar{\nu}})$,
where $\Delta$ is the distance to the critical point, and the critical
exponents $\beta$ and $\bar{\nu}$ depend on the degree exponent $\gamma$ and
are defined only for $\gamma>3$ \footnote{Being zero the critical
  temperature for $\gamma \leq 3$, FSS has no meaning in this
  region.}. In the context of nonequilibrium phase transitions,
however, the situation is not so clear. In the case of the contact
process (CP) in SF networks \cite{castellano06:_non_mean} Hong
\textit{et al.}  \cite{hong07} proposed a FSS for the average density
of particles in surviving runs scaling as $\rho_s(\Delta, N) =
N^{-\beta/\bar{\nu}}f(\Delta N^{1/\bar{\nu}})$, with
$\beta=1/(\gamma-2)$, $\bar{\nu}=(\gamma-1)/(\gamma-2)$ for $2 <
\gamma \leq3$, and $\beta=1$, $\bar{\nu}=2$ for $\gamma>3$. This
theory, however, is still under debate~\cite{ha07, castellano07:_reply},
since numerical simulations in random neighbors SF networks yield
incompatible results \cite{castellano07:_reply}.
For other kinds of FSS, see also Ref.~\cite{igloi}.

In this Letter we provide a step forward in the understanding of
finite size behavior in dynamical processes on SF networks by
showing that two different scaling scenarios may occur.  A first class
of processes exhibits finite size effects depending exclusively on the
network size $N$. On the other hand, a second class of processes
displays anomalous finite size effects, in the sense that their
properties depend explicitly and independently not only on the number
of nodes $N$, but also on the upper cutoff $k_c$ of the degree
distribution \cite{dorogorev}. Hints to this fact can be found in
previous works in which explicit solutions for dynamical models on SF
networks with infinite size but finite cutoff were obtained
\cite{pvbrief}. In this case, an explicit dependence on $k_c$ was
found, leading to a radically different behavior from that obtained
for truly infinite SF networks, which then do not correspond to the
analytical continuation to infinite $k_c$. 
The true thermodynamic limit of SF networks
corresponds to the double limit $N\to\infty$ and $k_c\to\infty$, since
keeping $k_c$ fixed leads to a finite degree second moment even for
$\gamma<3$, and to a non SF network.  The natural way to take this
double limit is to allow $k_c$ to diverge with $N$ but not faster than
$N^{1/2}$ if the network is to be uncorrelated \cite{mariancutofss}.
Hence any choice $k_c(N) \sim N^{1/\omega}$ with $2 \leq \omega <
\infty$ is legitimate and the way to reach the thermodynamic limit is
not unique.  Therefore, in systems with anomalous finite size effects,
depending on the way $k_c$ scales with system size, the final behavior
as a function of $N$ is modified, and will depend on the particular
value of $\omega$ chosen.
The heterogeneous mean-field (MF) theory for CP turns out to display
anomalous scaling, therefore our findings clarify the discrepancies
between MF theory and random-neighbor simulations recently reported in the
literature~\cite{castellano07:_reply}.
More generally, our results
indicate that the FSS analysis with the standard form
(i.e. as a function of $N$ alone) used in several recent works
for SF quenched networks should be reconsidered.

\begin{figure}
\includegraphics[width=75mm]{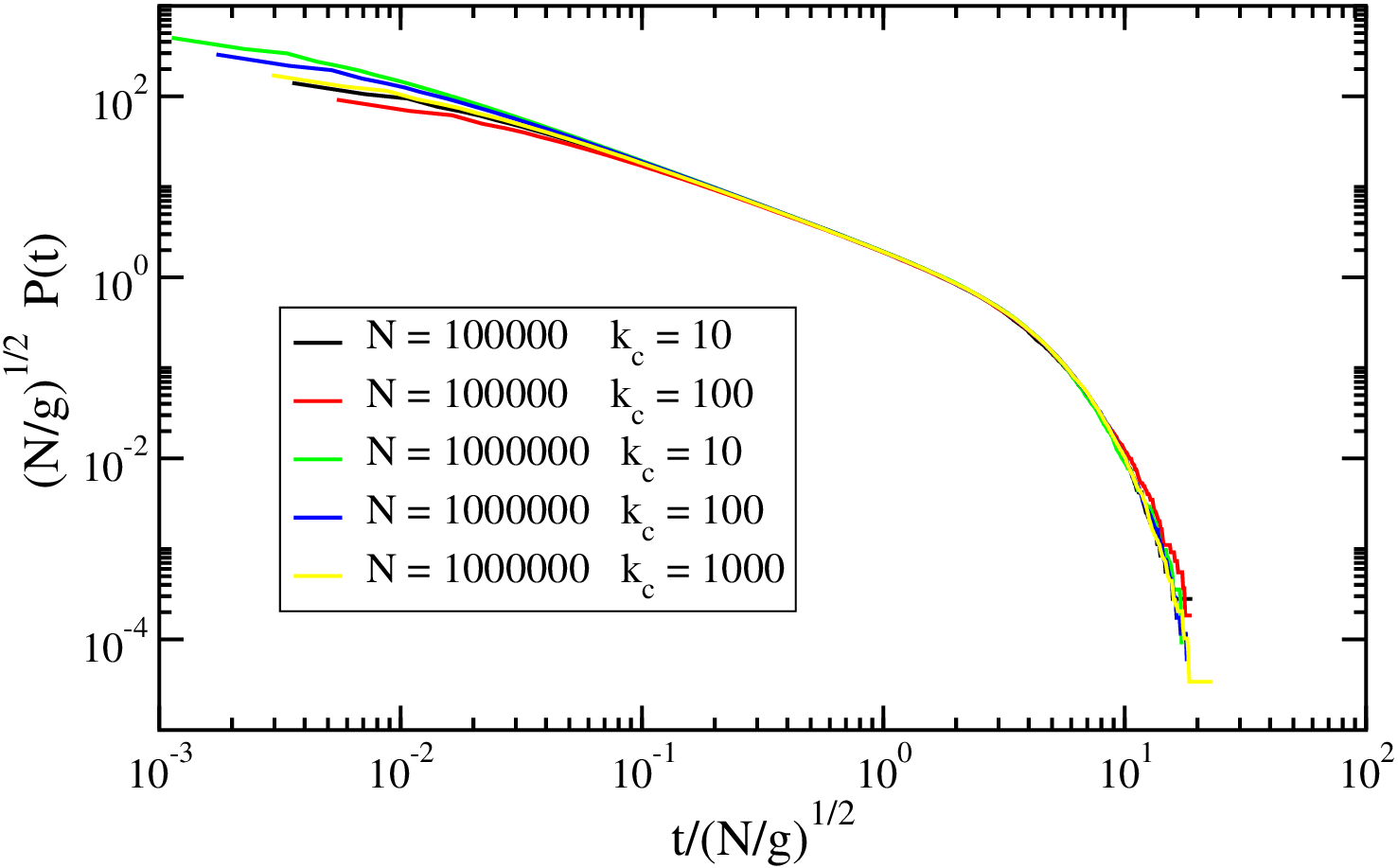}
\caption{Scaling plot of the survival probability of the CP on random
  neighbors SF networks with $\gamma=2.5$ and $k_c \le
  N^{1/2}$. Minimum degree $k_{\mathrm{min}}=2$.
  Numerically we obtain $\delta=1$.}
\label{Poft}
\end{figure}

We start our case by considering the contact process~\cite{marro99}
on heterogeneous networks, which is defined as
follows \cite{castellano06:_non_mean}. An initial fraction $\rho_0$ of
vertices is randomly chosen and occupied by a particle.  Dynamics
evolves in continuous time by the following stochastic processes:
Particles in vertices of degree $k$ create offsprings into their
nearest neighbors at rate $\lambda/k$, independently of the degree
$k'$ of the nearest neighbors. At the same time, particles disappear
at a unit rate~\footnote{For a discrete time implementation of this
  model, see Ref.~\cite{castellano06:_non_mean}.}.  This model
undergoes a continuous transition, located at a critical point
$\lambda_c$, separating an absorbing phase from an active one with
everlasting activity~\cite{marro99}.  The critical properties of CP on
uncorrelated SF networks have been studied by means of heterogenous
MF theory ~\cite{castellano06:_non_mean} and a FSS theory
\cite{hong07}. Heterogeneous MF theory predicts in the thermodynamic
limit a critical point $\lambda_c=1$, independent of the network
topology, a stationary particle density in the active phase given by
$\rho \sim \Delta^\beta$, with $\Delta=\lambda-\lambda_c$ and
$\beta=1/(\gamma-2)$, and a density decay at criticality $\rho(t) \sim
t^{-\theta}$, with $\theta=1/(\gamma-2)$. On the other hand,
Ref.~\cite{hong07} assumes a standard FSS, depending only on $N$,
which yields the particle density in surviving runs at criticality
\begin{equation}
  \rho_s \sim N^{-1/(\gamma-1)},
  \label{eq:2}
\end{equation}
for any $k_c > N^{1/\gamma}$.  Apart from the unsolved question of the
validity of heterogeneous MF theory for CP on uncorrelated SF networks
\cite{castellano06:_non_mean, ha07, castellano07:_reply}, there is
surprising evidence of a disagreement between the MF FSS exponents and
simulations on a random neighbor version of SF networks
\cite{castellano07:_reply}.  In order to gain an understanding of
finite size effects in the CP, we propose to focus on spreading
experiments \cite{marro99}, i.e. in simulations starting with a single
active site ($\rho(t=0)=\rho_0=1/N$) in which activity is followed until
the systems decays to the absorbing state. In such experiments it is
customary to measure the survival probability $P(t)$, defined as the
probability that activity survives up to time $t$.  At the critical
point this quantity scales as~\cite{marro99}
\begin{equation}
  P(t) = t^{-\delta} f(t/t_c),
  \label{pt}
\end{equation}
where the scaling function $f(x)$ is constant for small values of the
argument and cutoff exponentially for $x \gg 1$, and $t_c$ is a
characteristic time, depending usually on the size of the system.
Standard homogeneous MF FSS theory predicts that at criticality the
temporal cutoff scales as $t_c \sim N^{1/2}$~\cite{marro99}.  To gain
insight into the spreading experiment, we can map the dynamics at the MF
level onto an effective one-dimensional diffusion problem.  When a spreading
experiment for the CP is performed the number of active sites $n(t)$
starts at 1 and at each time step can grow by 1, decrease by 1 or
remain constant.  Defining the density of active sites
$\rho(t)=n(t)/N$, the diffusion process is defined in uncorrelated
networks at criticality by the transition rates
\begin{equation}
  \omega_{\rho \to \rho - \frac{1}{N}} = \rho(t), \; \omega_{\rho
    \to \rho + \frac{1}{N}} = \rho(t) \sum_k \frac{P(k) k}{\langle k
    \rangle} [1-\rho_k(t)], \label{eq:1} 
\end{equation}
where the last term in Eq.~(\ref{eq:1}) represents the probability
that a random neighbor is empty, and $\rho_k(t)$ is the density of
active sites restricted to nodes of degree $k$.  Clearly $\rho(t)=\sum_k
\rho_k(t) P(k)$.  The equation of motion for the $\rho_k$ at criticality
is~\cite{castellano06:_non_mean} 
\begin{equation}
  \partial_t  \rho_k(t) =
- \rho_k(t) + \frac{k}{\avk{k}} [1-\rho_k(t)] \rho(t).
\label{rho_k}
\end{equation}
Since all $\rho_k$ are expected to change in time slower than
exponentially, it is correct to perform a quasi-static
approximation~\cite{michelediffusion} and set the l.h.s. of
Eq.~(\ref{rho_k}) to zero, so that
\be 
\rho_k(t) \simeq \frac{k \rho(t) / \avk{k}}{1+ k \rho(t) / \avk{k}}
\label{rhoktot}
\ee at sufficiently large times.  Provided $\rho(t) \ll \avk{k}/k$,
which is always true in spreading experiments, the $k$-dependent term
in the denominator can be neglected, i.e.  $\rho_k(t) \simeq k \rho(t)
/ \avk{k}$.  This relation is very well verified numerically.
Inserting it into the transition rates yields
\begin{equation}
  \omega_{\rho \to \rho -\frac{1}{N}}  = \rho(t), \quad
  \omega_{\rho \to \rho+\frac{1}{N}} = \rho(t)  \left[1-g \rho(t)
  \right], \label{tr}
\end{equation}
where $g=\langle k^2 \rangle/\langle k \rangle^2$.
Hence the density $\rho(t)$ of active states performs a biased,
one-dimensional random walk, and
the distribution Eq.~(\ref{pt}) is the solution of the first passage time
problem in $\rho=0$, starting from the initial condition $\rho_0=1/N$.
Using the transitions rates~(\ref{tr}) it is possible to measure
the time $t_c$ with arbitrary precision in trivial simulations of the 
random walk process. 
It turns out that this characteristic time depends explicitly on
both the network size $N$ and the network cut-off $k_c$, through the
form
\begin{equation}
  t_c \sim \left(N / g \right)^{1/2}.
\label{tc}
\end{equation}
For $\gamma>3$, (non SF networks), $g$ is a constant and
Eq.~(\ref{tc}) coincides with the usual MF result.  For $\gamma
\leq 3$, on the other hand, $g$ diverges with the upper cutoff $k_c$ of
the degree distribution as $k_c^{3-\gamma}$, giving an anomalous form
of FSS, which separately and explicitly involves the
system size $N$ and the cutoff $k_c$, i.e.
\begin{equation}
t_c \sim N^{1/2} k_c^{(\gamma-3)/2}.
\end{equation}
We have verified numerically, by performing spreading experiments on a
random neighbors version of the uncorrelated configuration model
\cite{ucmmodel} with $k_c<N^{1/2}$, that the surviving probability of
the CP at criticality is well described by Eq.~(\ref{pt}) with a
characteristic time $t_c$ given by Eq.~(\ref{tc}), see
Fig.~\ref{Poft}.
As we have discussed above, the cutoff $k_c$ is not
fixed, and its divergence with network size can be chosen with a
degree of arbitrariness.  Imposing therefore $k_c(N) \sim
N^{1/\omega}$ and expressing $t_c$ only as a function of $N$, this
arbitrariness results in a dependence on the exponent $\omega$,
\begin{equation}
t_c \sim N^{[1+(\gamma-3)/ \omega ]/2}.
\end{equation}
This anomalous dependence of $t_c$ on both $N$ and $k_c$ translates
into a density of active sites in surviving runs $\rho_s$ (the usual
observable for FSS in non-equilibrium absorbing-state phase
transitions) depending also separately on $N$ and $k_c$.  To determine
its detailed expression, it is crucial to realize that the solution of
the MF Eq.~(\ref{rho_k}) for $\rho(t)$ at criticality shows a
crossover at a temporal scale $t^* \sim k_c^{\gamma-2}$, in full
analogy with the crossover occurring for diffusion-annihilation
processes in SF networks~\cite{michelediffusion}.  The origin of the
crossover can be traced back to Eq.~(\ref{rhoktot}).  When there is a
finite initial density $\rho_0$ of active sites, the $k-$dependent
term in the denominator of Eq.~(\ref{rhoktot}) cannot be neglected
and hence, for $t_\times < t < t^*$, the density
decays as if the network size were effectively infinite, so that
$\rho(t) \sim t^{-1/(\gamma-2)}$~\cite{castellano06:_non_mean}.
$t_\times$ is a microscopic time scale, independent of $N$ and $k_c$.
As the density decreases, for times larger than $t^*$, such that
$k_c \rho(t^*) =  \avk{k}$, the denominator of Eq.~(\ref{rhoktot}) does
not effectively depend on $k$ anymore so that
for $t^* < t < t_c$, $\rho(t) \sim (gt)^{-1}$ for any $\gamma$,
yielding an exponent $\theta=1$.  The asymptotic value in
surviving runs $\rho_s$ is reached for times larger than $t_c$, so
$\rho_s \sim \rho(t_c)$. Since $t_c>t^*$ for $\gamma<3$ and
$\omega>2$, one has at the critical point
\begin{equation}
  \rho_s \sim  \frac{1}{g t_c} \sim N^{-1/2} k_c^{(\gamma-3)/2}
\sim N^{-[1-(\gamma-3)/\omega ]/2}.
\label{rhos}
\end{equation}
Eq.~(\ref{rhos}) shows that not only $t_c$ but also other observables
acquire an anomalous $k_c-$dependence on finite SF networks, which
translates into values of the exponent ratio $\beta/\bar{\nu}$
depending on the particular $\omega$ considered.  This result accounts for
the numerical difference in the $\beta/\bar{\nu}$ value found in random
neighbor simulations with different values of
$\omega$~\cite{castellano07:_reply}, which can be better
fitted~\footnote{Except for $\gamma \to 2$. Data in
Ref.~\cite{castellano07:_reply} are probably not asymptotic in such limit.}
by means of Eq.~(\ref{rhos}), whose decay exponent is smaller than the one
proposed in Refs.~\cite{ha07,hong07}, namely Eq.~(\ref{eq:2}).
It is also worth stressing that, since $t_c>t^*$, the true
asymptotic MF value for the temporal exponent $\theta$ describing the
decay of density is $1$ for any $\gamma$ and any $N<\infty$, and not
the exponent $1/(\gamma-2)$ predicted for $\gamma<3$ by the
heterogeneous MF treatment for strictly infinite system
size~\cite{ha07,hong07}.

\begin{figure}
\includegraphics[width=75mm]{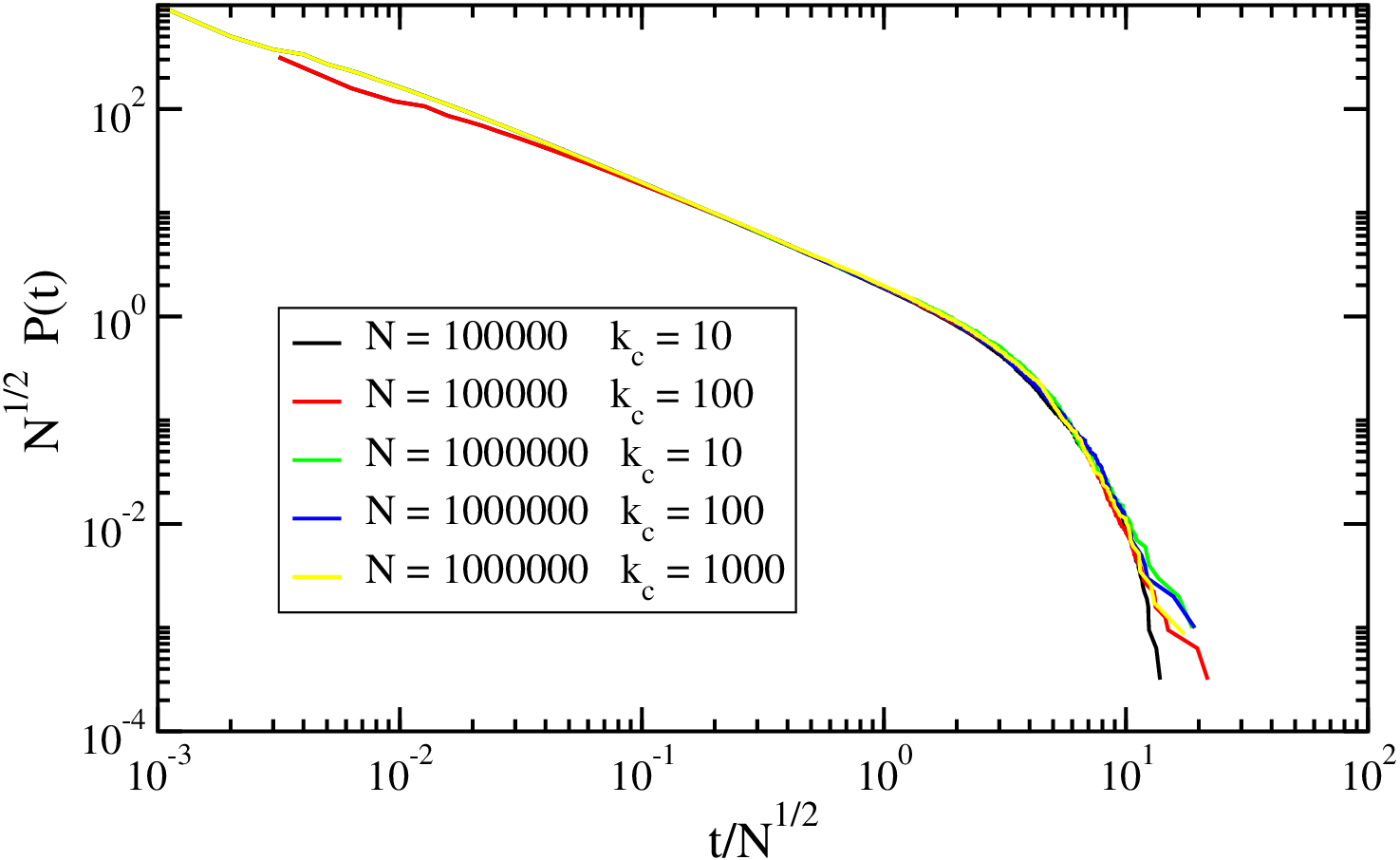}
\caption{Scaling plot of the survival probability of the modified CP
  in Ref.~\cite{giuranic05} with $\mu=1$, on random neighbors SF
  networks with $\gamma=2.5$ and $k_c \le N^{1/2}$. Minimum degree
  $k_{\mathrm{min}}=2$.}
\label{Poft_CP-k}
\end{figure}

This finite size separate dependence of scaling time at criticality on
both the system size and the upper cutoff is not in fact a peculiarity
of the CP in MF, but is actually found also in other processes taking
place on SF networks. Sood and Redner~\cite{Sood05} have recently
computed the time needed by the voter model dynamics taking place on
heterogeneous networks to reach a full consensus (ordered) state,
which is given by $\tau \sim N/g$, and hence depends on both $N$ and
$k_c$ for $\gamma \leq 3$. On the other hand, the distinct dependence
on $N$ and $k_c$ is not the only possibility: There are systems for
which the size scaling on heterogeneous networks is not anomalous and
depends exclusively on $N$.  Two examples are provided by the
``link-update''~\cite{Suchecki05} and
``reverse''~\cite{castellano:114} versions of the voter model.  In the
first case the time to reach full consensus is proportional to
$N$~\cite{castellano:114}, with no dependence on network features
whatsoever.  For the reverse model instead such time is proportional
to $N \avk{k} \avk{k^{-1}}$~\cite{castellano:114}.  For all $\gamma >
2$ these moments of the degree distribution are well-behaved and no
dependence on $k_c$ arises. In this same spirit, it is possible to
devise stochastic systems with continuous phase transitions that show
standard finite size effects, depending only on the network
size. Consider, for example, the variation of the CP proposed in
Ref.~\cite{giuranic05} (see also \cite{igloi}) in which particles in
vertices of degree $k$ create offsprings in nearest neighbors of degree
$k'$ at rate $\lambda (k k')^{-\mu}$, while particles still disappear
at unit rate. For $\mu=1$, the heterogeneous MF theory for this model
takes, in uncorrelated networks, the form
\begin{equation}
  \partial_t  \rho_k(t) =
  - \rho_k(t) + \frac{\lambda \rho(t)}{\avk{k}} [1-\rho_k(t)],
\end{equation}
which has a threshold $\lambda_c = \avk{k}$, and a
stationary particle density in the active phase $\rho \sim \lambda -
\lambda_c$. It is easy to see that the random walk mapping of
spreading experiments at criticality in this modified CP model is
defined by the transition rates in Eq.~(\ref{tr}) with $g=1$, which
correspond to a characteristic time scaling as $t_c \sim N^{1/2}$,
completely independent of the network cutoff $k_c$. This result is
fully confirmed by numerical simulations, see Fig.~\ref{Poft_CP-k}.

\begin{figure}
\includegraphics[width=7.5cm]{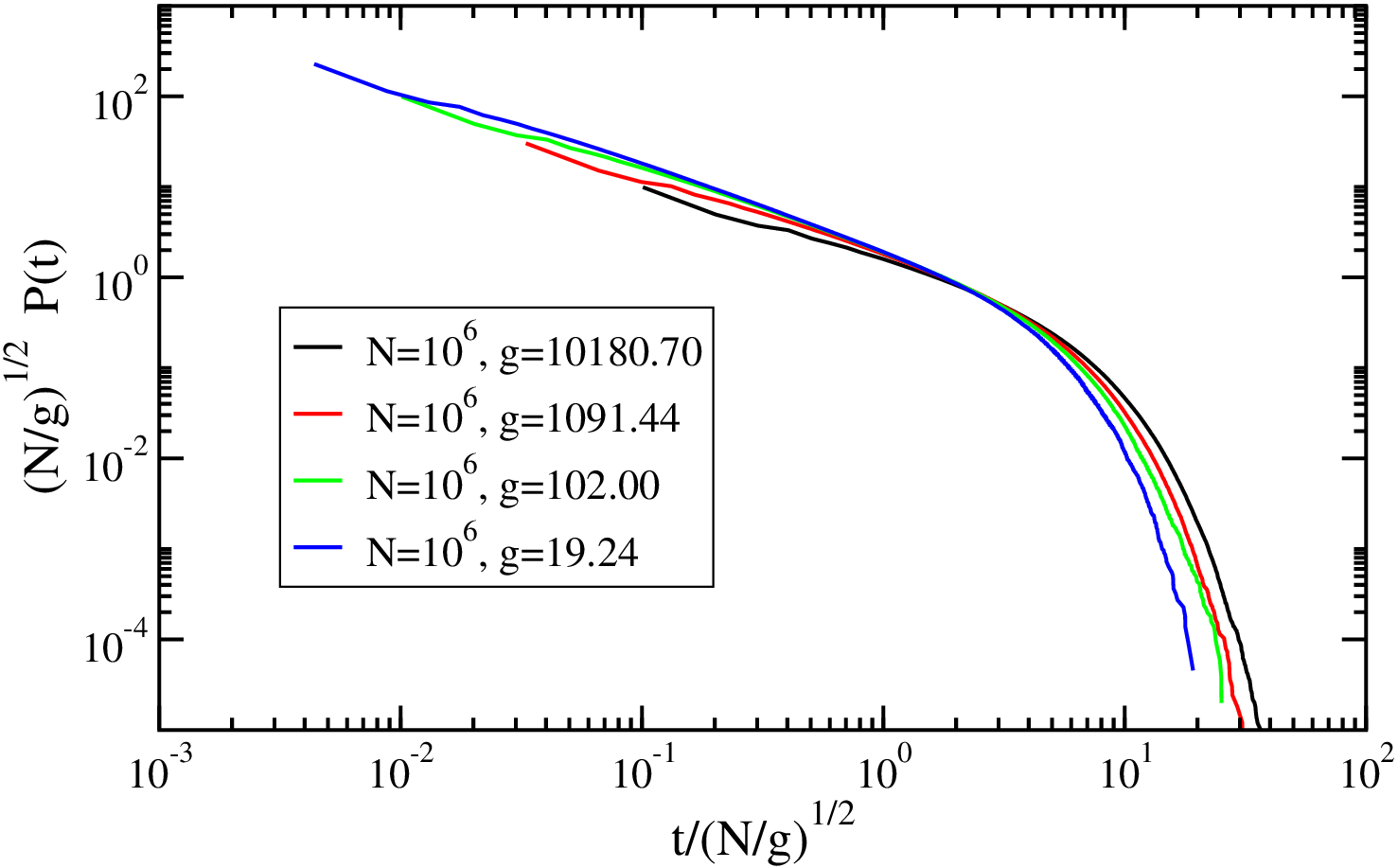}
\caption{Scaling plot of the survival probability of the CP on random
  neighbors SF networks with $\gamma=2.5$ and $k_c =
  N^{1/(\gamma-1)}$. Minimum degree $k_{\mathrm{min}}=2$. Notice that,
  for $\omega<2$, $g$ displays very large sample to sample fluctuations.}
\label{Poft2}
\end{figure}

Finally, let us discuss what happens in the standard CP when the cutoff
scales as $k_c \sim N^{1/\omega}$ but with $\omega<2$.  This case
includes the so-called natural cutoff $\omega=\gamma-1$ (for $\gamma<3$)
that occurs in the configuration model if one does not impose any
constraint on the maximum degree of the
vertices~\cite{dorogorev,mariancutofss}. 
In such a case, the presence of very large hubs invalidates the
derivation of the Eq. (7). Figure~\ref{Poft2} shows that this is
in fact the case for
$\gamma<3$ and $\omega=\gamma-1<2$, and it is then the proof that
for random neighbor networks with $k_c > N^{1/2}$ a different scaling
form of $t_c$ must be considered.

In conclusion, the evidence presented in this work leads to
the conclusion that processes taking place on SF networks may be divided
in two classes. In the first class, finite size effects and FSS depend only
on the number of nodes $N$. This class encompasses ``reverse'' and
``link-update'' voter dynamics as well as models showing continuous
phase transitions, such as the Ising model for $\gamma>3$, and
modified versions of the CP.  Other models, on the other hand, show an
additional explicit dependence on the upper cutoff of the degree
distribution.  In this last case the thermodynamic limit of an
infinitely large network can be reached in different ways, depending
on how $k_c$ is chosen to scale with $N$, and different possible
routes to the thermodynamic are thus possible.
Understanding what physical ingredients select which class
a model belongs to is a challenging task for future work.

\begin{acknowledgments}
  R.P.-S. acknowledges financial support from the Spanish MEC (FEDER),
  under projects No. FIS2004-05923-C02-01 and
  No. FIS2007-66485-C02-01.
\end{acknowledgments}


\begin{thebibliography}{10}

\bibitem{goldenfeld}
N. Goldenfeld, {\em Lecture notes on phase transitions and the renormalization
  group}, {\em Frontiers in Physics} (Addison-Wesley, Massachusetts, 1992).

\bibitem{privman90}
V. Privman, {\em Finite Size Scaling and Numerical Simulation of Statistical
  Systems} (World Scientific, Singapore, 1990).

\bibitem{barabasi02}
R. Albert and A.-L. Barab{\'a}si, Rev. Mod. Phys. {\bf 74},  47  (2002).

\bibitem{mendesbook}
S.~N. Dorogovtsev and J.~F.~F. Mendes, {\em Evolution of networks: From
  biological nets to the {I}nternet and {WWW}} (Oxford University Press,
  Oxford, 2003).

\bibitem{watts98}
D.~J. Watts and S.~H. Strogatz, Nature {\bf 393},  440  (1998).

\bibitem{dorogovtsev07:_critic_phenom}
S. Dorogovtsev, A. Goltsev, and J. Mendes, Critical phenomena in complex
  networks, 2007, e-print arXiv:0705.0010v2.

\bibitem{castellano06:_non_mean}
C. Castellano and R. Pastor-Satorras, Phys. Rev. Lett. {\bf 96},  038701
  (2006).

\bibitem{hong07}
H. Hong, M. Ha, and H. Park, Phys. Rev. Lett. {\bf 98},  258701 (2007);

\bibitem{ha07} M. Ha, H. Hong, and H. Park, Phys. Rev. Lett. {\bf 98},
029801  (2007).

\bibitem{castellano07:_reply}
C. Castellano and R. Pastor-Satorras, Phys. Rev. Lett. {\bf 98},  029802
  (2007).

\bibitem{igloi}
  M. Karsai, R. Juhasz, and F. Igloi, Phys. Rev. E \textbf{73}, 036116
  (2006).

\bibitem{dorogorev}
S.~N. Dorogovtsev and J.~F.~F. Mendes, Adv. Phys. {\bf 51},  1079  (2002).

\bibitem{pvbrief}
R. Pastor-Satorras and A. Vespignani, Phys. Rev. E {\bf 65},  035108(R) (2002).

\bibitem{mariancutofss}
M. Bogu{\~n}{\'a}, R. Pastor-Satorras, and A. Vespignani, Euro. Phys. J. B {\bf
  38},  205  (2004); Z. Burda, A. Krzywicki, Phys. Rev. E {\bf 67},
046118 (2003).


\bibitem{marro99}
J. Marro and R. Dickman, {\em Nonequilibrium phase transitions in lattice
  models} (Cambridge University Press, Cambridge, 1999).

\bibitem{michelediffusion}
M. Catanzaro, M. Bogu{\~n}{\'a}, and R. Pastor-Satorras, Phys. Rev. E {\bf 71},
   056104  (2005).

\bibitem{ucmmodel}
M. Catanzaro, M. Bogu{\~n}{\'a}, and R. Pastor-Satorras, Phys. Rev. E {\bf 71},
   027103  (2005).



\bibitem{Sood05}
V. Sood and S. Redner, Phys. Rev. Lett. {\bf 94},  178701  (2005).

\bibitem{Suchecki05}
K. Suchecki, V.~M. Eguiluz, and M. {San Miguel}, Europhys. Lett. {\bf 69},  228
   (2005).

\bibitem{castellano:114}
C. Castellano,  AIP Conf. Proc. \textbf{779}, 114 (2005).

\bibitem{giuranic05}
  C.V~Giuraniuc \textit{et al.}, Phys. Rev. Lett. \textbf{95}, 098701
  (2005). 

\end{thebibliography}
\end{document}